\documentclass[a4paper]{article}

\usepackage[english]{babel}
\usepackage[utf8x]{inputenc}
\usepackage[T1]{fontenc}

\usepackage[a4paper,top=3cm,bottom=2cm,left=3cm,right=3cm,marginparwidth=1.75cm]{geometry} 
\usepackage{amsmath}
\usepackage{graphicx}
\usepackage[colorinlistoftodos]{todonotes}
\usepackage[colorlinks=true, allcolors=blue]{hyperref}

\DeclareMathOperator{\sgn}{sgn}
\DeclareMathOperator{\tr}{tr}

\newcommand{\units}[1]{{\,\rm #1}}

\newcommand{\eq}[1]{Eq.\,(\ref{#1})}
\newcommand{\Eq}[1]{Equation\,(\ref{#1})}
\newcommand{\eqs}[2]{Eqs.\,(\ref{#1},\ref{#2})}
\newcommand{\eqS}[3]{Eqs.\,(\ref{#1},\ref{#2},\ref{#3})}
\newcommand{\eqss}[2]{Eqs.\,(\ref{#1}--\ref{#2})}

\usepackage{authblk}

\title{A Unique Analytical Solution of the White Matter Standard Model using Linear and Planar Encodings}
\author[1,2,3]{Marco Reisert}
\author[1,3]{Valerij G.\ Kiselev}
\author[1,3,4]{Bibek Dhital}

\affil[1]{Medical Center, Faculty of Medicine, University
Freiburg, Germany}
\affil[2]{Department of Functional and Stereotactic Neurosurgery, Freiburg}
\affil[3]{Department of Medical Physics, Freiburg}
\affil[4]{Department of Neurophysics, Max Planck Institute for Human Cognition and Brain Sciences, Leipzig, Germany}

\begin{document}
\maketitle

\begin{abstract}
Diffusion-weighted magnetic resonance imaging in brain white matter probes tissue microstructure and allows for the estimation of compartmental diffusion parameters. Recently, it became apparent that traditional single-direction diffusion encodings are not fully sufficient to resolve the white matter compartmental diffusivities. 
Multiple diffusion encodings have been suggested to make the problem less 
ambiguous, however, it still remained unclear whether such protocols would completely solve the problem.
Here, we constructively prove that a combination of linear and planar diffusion 
encodings is enough to determine the parameters of the three compartment white matter model. 
\end{abstract}

\newcommand{\Di}{D_i}
\newcommand{\De}{\Delta_e}
\newcommand{\Dr}{D_e}
\newcommand{\Df}{D_\text{f}}
\newcommand{\vi}{v_\text{i}}
\newcommand{\ve}{v_\text{e}}
\newcommand{\vf}{v_\text{f}}
\newcommand{\Mp}{M_\text{pla}}
\newcommand{\Ml}{M_\text{lin}}
\newcommand{\Ms}{S_\text{sph}}
\newcommand{\mv}[1]{\mathbf{#1}}

\newcommand{\vk}[1]{{\color{red} VK: #1}}

\pagestyle{myheadings}
\markright{Submitted to Magnetic Resonance in Medicine (MRM)}

\section{Introduction}

For a long time, attempts to multi-compartment modeling in brain white matter (WM) with simple single diffusion encodings \cite{fieremans2011white,zhang2012noddi,novikov2018rotationally,reisert2017disentangling} 
led to ambiguous results \cite{jelescu2016degeneracy,novikov2018rotationally}. 
For example, it was argued in 
\cite{fieremans2011white} that intra axonal diffusion is substantially smaller than 
extra axonal diffusion along the axons, while others argued for the opposite
\cite{zhang2012noddi,dhital2017absence}. 
%\vk{add NYU, remove NODDI?}.
Multiple diffusion encodings offer 
substantially more information than ordinary single diffusion encoding schemes \cite{jespersen2013orientationally,westin2014measurement}.
However, most efforts in understanding the additional information gained by
such methods were focused on dispersed single-compartment systems thus revealing apparent measures like eccentricity, microscopic and fractional anisotropy \cite{jespersen2013orientationally,westin2014measurement,szczepankiewicz2015quantification}.

Recent studies have investigated the benefits of using  multiple diffusions encodings to resolve white matter compartmental parameters \cite{lampinen2017neurite, dhital2018}.
For example, spherical diffusion encodings \cite{dhital2017absence} shows very low kurtosis in white and gray matter, which gives rise to the assumption that traces of the tissue compartments are similar. 
In \cite{fieremans2018} an additional spherical encodings were used to stabilize fits and release constraints. 
Or, in \cite{coelo2017,reisert2018} a combination of linear and planar encodings was used with the same intention.
Thus, the question arises, what kind of protocol is sufficient to solve the problem 
uniquely? 
This short note contributes to the answer of this question.
%\vk{``contributes" too weak. }

We will show that a combination of linear and planar encodings is indeed enough to provide 
a unique solution of the full 3-compartment model of brain white matter using $\mathcal{O}(b^2)$ measurements.
The key ingredient of the approach is that a combination of linear and planar measurements provide
a direct estimate of the mesoscopic orientation dispersion, without relying on any other concurrent estimates. 
We further discuss an inherent model property, that, under special conditions,
this solution still shows an ambiguity.
Finally, we demonstrate by a few counterexamples the inadequacy of linear and spherical encoding to resolve the problem ( taking only $\mathcal{O}(b^2)$ coefficients).
%\vk{do it analytically.}

\section{The White Matter Model}
\label{sec:model}
We follow the standard tissue model as proposed in \cite{novikov2018rotationally,reisert2017disentangling}. 
In contrast to \cite{zhang2012noddi}, in this model both intra and extra-axonal compartments undergo the same convolution with the mesoscopic orientation
distribution. 
%This not only allows for the disentanglement of micro and mesoscopic contributions, but is  also  physically reasonable assumption. 
%\vk{as discussed below?}
%\vk{the latter is not self-evident. Let us discuss.} 
For a general encoding matrix $\mv B$ the signal for this model looks as follows 
\begin{eqnarray}
S(\mv B) &=& \int_{\mv n \in S_2} d^2 \mv n\  M(\mv n, \mv B) f(\mv n) \\
&=& \int_{\mv n \in S_2} d^2 \mv n \left( v_i e^{-\tr(\mv B \mv D_i^{\mv n})} + v_e e^{-\tr(\mv B \mv D_e^{\mv n})} + \vf e^{-\tr(\mv B)\Df} \right) f(\mv n)
\label{model}
\end{eqnarray}
where $f(\mv n)$ is an arbitrary, normalized orientation distribution function and 
$M(\mv n, \mv B)$ the axially symmetric, multi-exponential microstructural model 
with symmetry axis $\mv n$. The diffusion tensor of intra- and extra-axonal fractions are
parametrized as 
\[
\mv D_i^{\mv n} = \mv n \mv n^T \Di ,\  \mv D_e^{\mv n} = \mv n \mv n^T \De + \mv I_3 \Dr
\]

%and  spherical encodings $\mv B_\text{sph} = b \mv I_3/3 $.
We now focus on linear encoding $\mv B_\text{lin} = b \mv q \mv q^T$ and planar encoding $\mv B_\text{pla} = b (\mv I_3 - \mv q \mv q^T)/2$, where $\mv q$ is the diffusion 
gradient direction of modulus one and the b-value $b$ is defined as the trace of the b-matrix. 
Rewriting the microstructural model in terms of the cosine $t = \mv q^T\mv n$ between encoding direction
and axon orientation gives, 
\begin{eqnarray}
\Ml(t,b) &=& \vi e^{-b \Di t^2} +  \ve e^{-b \De t^2 - b\Dr} +  \vf e^{-b\Df} 
\label{Mlin}\\
\Mp(t,b) &=& \vi e^{-b \Di (1-t^2)/2} +  \ve e^{-b \De (1-t^2)/2 - b\Dr} +  \vf e^{-b\Df} 
\label{Mpla}
%\Ms(3b) &=& \vi e^{-b \Di} + \ve e^{-b (\De+3\Dr)} + \vf e^{-3\Df b}
\end{eqnarray}
In this formulation, the convolution with the mesostructural orientation distribution $f(\mv n)$ takes the form %\vk{meso- and micro not defined. }
\begin{eqnarray}
S_\alpha(\mv q,b) &=& \int_{\mv n\in S_2} d^2\mv n \ f(\mv n)\ M_\alpha(\mv q^T\mv n,b)
\label{S=fnMq}
\end{eqnarray}
where $\alpha = \text{linear}$ or $\alpha = \text{planar}$ depending on the gradient waveform. 
Note that $S_\alpha(\mv q,b)$ is normalized in the sense $S_\alpha(\mv q,0)=1$.

The key to decouple micro and mesostructural contribution is 
to work in the domain of spherical harmonics.
The spherical convolution turns out to be a product of the two spherical harmonic representations, $f_{l,m}$ and $M_\alpha^{l}(b)$,  of 
$f(\mv n)$ and $M_\alpha(t,b)$, respectively.
\begin{equation}
S_\alpha^{l,m}(b)  = f_{l,m}\ M_\alpha^{l}(b) \,.
\label{S=fM}
\end{equation}

We used here in semi-Schmidt normalization\footnote{
In this normalization $\sum_{m=-l}^l |Y_l^m(\mv n)|^2 = 1$ and 
$\int d^2 \mv n\, Y_l^m(\mv n) Y_{l'}^{m'}(\mv n)^* = \frac{4\pi}{2l+1}\delta_{l,l'}\delta_{m,m'}$
and $Y_l^0(\mv n) = P_l(\cos\theta)$, where 
$Y_l^m$ are the spherical harmonics and $P_l$ the Legendre polynomials and $\theta$ the polar angle of $\mv n$. The axial symmetry implies that the spherical harmonics expansion of $M_\alpha(\mv q^T\mv n,b)$ contains only components with $l=0$, $M_\alpha(\mv q^T\mv n,b)=\sum_l \frac{2l+1}{4\pi}M^l_\alpha(b)P_l(\mv q^T\mv n)$. }  as in \cite{reisert2017disentangling}. 
%\vk{move the axial symmetry in the main text. }
The signal is characterized by a set of quantities that are rotationally invariant for any signal-generating tissue. 
\begin{equation}
S_\alpha^{l}(b)  
=  \sqrt{\sum_{m=-l}^l |S_\alpha^{l,m}(b)|^2} 
= f_l\ |M_\alpha^{l}(b)| 
\label{eq:powers}
\end{equation}
Here $f_l = \sqrt{\sum_m | f_{l,m}|^2}>0$ is the rotation invariant mesoscopic dispersion.
For both linear and planar encodings, we define the moments 
%\vk{to mention: (i) we define the projections as $\int d^2 \mv n Y_{l,m}(\mv n)^*$ and (ii) mention normalization on $b=0$. }
\begin{align}
W_\alpha^{l,k} &:= \frac{1}{4\pi}\left. \frac{d^k}{db^k} \right|_{b=0} S_\alpha^{l}(b)  \label{W=} \\
&= f_l \,\sgn (M_\alpha^{l}(0)) \left. \int_{-1}^1 \frac{dt}{2}\ P_l(t) \frac{d^k}{db^k} \right|_{b=0}  M_\alpha(t,b) \,,
\end{align}
where $\sgn(x)=x/|x|$ and we do not write the delta-functional term for $l\geq 2$, since $M_\alpha^{l}(b)$ have definite signs. 
This follows from their physical meaning of the signal from the idealized unidirectional fiber bundle, since diffusion is faster along such a bundle, $M^2_\text{lin}<0$ and $M^2_\text{pla}>0$, for all meaningful constellations of microstructural parameters. 
Introduction of these definite signs is sufficient to resolve the ambiguity borne by taking the square of \eq{eq:powers}, which is necessary to build rotation invariant quantities. 

%A similar ambiguity was first noticed when using kurtosis for the parameter determination in a simplified model of unidirectional neuronal fibers \cite{fieremans2011white} and was shown to persist in more involved methods for parameter estimation \cite{jelescu2016degeneracy,novikov2016mapping}. 

Note that the moments defined in \eq{W=} generalize the moments used by 
\cite{novikov2018rotationally}; for linear encoding 
$W_\text{lin}^{l,k} \propto M^{(2k),l} $ following definitions in
\cite{novikov2018rotationally} equations (15-18).

\subsection{Finding the solution}

The white matter model described above includes one known (the free water diffusivity, $\Df$) and five unknown scalar parameters: intra-axonal difusivity ($\Di$), extra-axonal radial diffusivity ($\Dr$), difference between extra-axonal parallel and radial diffusivity ($\De$), and the volume fractions ($\vi$, $\ve$, $\vf$) with the constraint $v_i+v_e+v_f=1$. The orientation distribution function $f(\mv n)$ contains an infinite set of coefficients. In this section we show that resolving the signal for both linear and planar encoding up to the order $b^2$ and $l=2$ enables unambiguous determination of the scalar parameters and the first non-trivial coefficient, $f_2$, of $f(\mv n)$.  

The corresponding moments are expressed via the model parameters as follows:
%% original equations:
% \begin{align} 
% W_\text{lin}^{0,1}&=- (\De \ve)/3 - \Dr \ve - (\Di \vi)/3 - \Df \vf \nonumber \\ 
% W_\text{lin}^{2,1}&=2 f_2 (\De \ve +  \Di \vi)/15 \nonumber \\ 
% W_\text{lin}^{0,2}&=(\De^2 \ve)/5 + \Dr^2 \ve + (\Di^2 \vi)/5 + \Df^2 \vf + (2 \De \Dr \ve)/3 \nonumber \\ 
% W_\text{lin}^{2,2}&=-f_2 \left[(4 \De^2 \ve)/35 + (4 \Di^2 \vi)/35 + (4 \De \Dr \ve)/15\right] \nonumber \\ 
% W_\text{pla}^{0,2}&=(2 \De^2 \ve)/15 + \Dr^2 \ve + (2 \Di^2 \vi)/15 + \Df^2 \vf + (2 \De \Dr \ve)/3 \nonumber \\
% W_\text{pla}^{2,2}&=-f_2 \left[(4 \De^2 \ve)/105 + (4 \Di^2 \vi)/105 + (2 \De \Dr \ve)/15\right] \nonumber 
% \end{align}
\begin{eqnarray} 
W_\text{lin}^{0,1}&=& - \frac{1}{3} \De \ve  - \Dr \ve - \frac{1}{3} \Di \vi  - \Df \vf  \label{weq1} \\ 
W_\text{lin}^{2,1}&=& \frac{2}{15} f_2 [\De \ve +  \Di \vi] \label{weq2} \\ 
W_\text{lin}^{0,2}&=& \frac{1}{5} \De^2 \ve  + \Dr^2 \ve + \frac{1}{5} \Di^2 \vi  + \Df^2 \vf +  \frac{2}{3} \De \Dr \ve   \label{weq3} \\ 
W_\text{lin}^{2,2}&=& -f_2 \left[ \frac{4}{35} \De^2 \ve  +  \frac{4}{35} \Di^2 \vi  +  \frac{4}{15} \De \Dr \ve \right]  \label{weq4} \\ 
W_\text{pla}^{0,2}&=& \frac{2}{15} \De^2 \ve  + \Dr^2 \ve +  \frac{2}{15} \Di^2 \vi  + \Df^2 \vf +  \frac{2}{3} \De \Dr \ve  \label{weq5} \\
W_\text{pla}^{2,2}&=& -f_2 \left[ \frac{4}{105} \De^2 \ve  +  \frac{4}{105} \Di^2 \vi  +  \frac{2}{15} \De \Dr \ve \right]  \label{weq6}
\end{eqnarray}
The calculations straightforwardly follow from, \eq{W=}.
% , with the substitution of $t^2$ and $t^4$ with their well known expressions in terms of Legendre polynomials, 
% \begin{align} 
% 1 &= P_0(t) \label{t0}  \\ 
% t^2 &= \frac{1}{3}[P_0(t) + 2P_2(t)] \label{t2} \\  
% t^4 &= \frac{1}{35}[7P_0(t)+20P_2(t)+8P_4(t)]\,. \label{t4} 
% \end{align}
%Contribution of $P_4(t)$ is omitted, since it is of the order $l=4$. 
Note the absence of the moments $W_\text{pla}^{l,1}$ - in this order (linear in $b$) measurements with any shape of $\mv B$ is equivalent to a set of single-direction measurements and thus do not add any extra information. 
For example, the signal obtained using the planar encoding in the $x,y$ plane is equivalent to the mean of signals encoded linearly in the $x$ and $y$ directions.
In particular,  
$W_\text{lin}^{0,1} =  W_\text{pla}^{0,1} \label{Wlin=Wpla}$ and 
$W_\text{lin}^{2,1} = 2 W_\text{pla}^{2,1} \label{Wlin=2Wpla}$,
which can be observed from the fact that $\Mp(t,b)$, \eq{Mpla}, can be obtained from $\Ml(t,b)$, \eq{Mlin}, by substituting $t^2$ with $(1-t^2)/2=[P_0(t)-P_2(t)]/3$. 
Complimentary information can be found in the second (or higher) order of $b$.

The dispersion parameter, $f_2$, can be easily found from \eq{weq3}-\eq{weq6}
\begin{eqnarray} 
f_2 =    -\frac{7}{4} \, \frac{W_\text{lin}^{2,2} - 2W_\text{pla}^{2,2}}{  W_\text{lin}^{0,2} - W_\text{pla}^{0,2} }
\label{f2=}
\end{eqnarray}
Note that the denominator is just the average eccentricity 
of the compartments \cite{jespersen2013orientationally} (or microstructural fractional anisotropy), namely
$W_\text{lin}^{0,2} - W_\text{pla}^{0,2} = 
 \De^2 \ve +  \Di^2 \vi$.

Finding other parameters is not so straightforward. 
Assuming $f_2$ is known, we define a set of auxiliary variables $x_i$ as follows
%% Original equation:
% \begin{eqnarray}
% x_1&=& (15 W_\text{lin}^{2,1})/(2 f_2)=\De \ve + \Di \vi\nonumber \\
% x_2&=& W_\text{lin}^{0,2} - W_\text{pla}^{0,2} =\De^2 \ve + \Di^2 \vi\nonumber \\
% x_3&=& - W_\text{lin}^{0,1} - (5 W_\text{lin}^{2,1})/(2 f_2)=\Dr \ve + \Df \vf\nonumber \\
% x_4&=& W_\text{lin}^{0,2} + W_\text{lin}^{2,2}/(4 f_2) + (9 W_\text{pla}^{2,2})/(2 f_2)=\Dr^2 \ve + \Df^2 \vf\nonumber \\
% x_5&=& (15 W_\text{lin}^{2,2})/(2 f_2) - (45 W_\text{pla}^{2,2})/(2 f_2)=\De \Dr \ve\nonumber 
% \end{eqnarray}
\begin{alignat}{3}
x_1&=  \frac{15}{2 f_2} W_\text{lin}^{2,1} &=\De \ve + \Di \vi \label{x1=} \\
x_2&= W_\text{lin}^{0,2} - W_\text{pla}^{0,2} &=\De^2 \ve + \Di^2 \vi \label{x2=} \\
x_3&= - W_\text{lin}^{0,1} -  \frac{5}{2 f_2} W_\text{lin}^{2,1} &=\Dr \ve + \Df \vf \label{x3=}  \\
x_4&= W_\text{lin}^{0,2} + \frac{1}{4 f_2}W_\text{lin}^{2,2} +  \frac{9}{2 f_2} W_\text{pla}^{2,2} &=\Dr^2 \ve + \Df^2 \vf \label{x4=} \\
x_5&=  \frac{15}{2 f_2} W_\text{lin}^{2,2}  -  \frac{45}{2 f_2} W_\text{pla}^{2,2} &=\De \Dr \ve \label{x5=} 
\end{alignat}
This system including the constraint on the compartment water fractions 
\begin{equation}
\vi+\ve+\vf=1  \label{volfrac}  
\end{equation}
defines all scalar parameters. 
% \subsubsection{Two Compartments}
% We first consider to the simplified model with no free water, $\vf=0$. The above equation system is overdefined for such a model; the parameters should be found by fitting \eq{x1=} -- \eq{x5=} to data with the constraint imposed by \eq{sum_v=1}. 
% \vk{do we really need to show a solution obtained by arbitrary dropping an equation? If so, one can also drop \eq{x2=}. }
% The solution is obvious
% \begin{eqnarray}
% \Dr = \frac{x_4}{x_3}\,,\ \ve = \frac{x_3^2}{x_4}\,,\ \De = \frac{x_5}{x_3}\, \\
% \Di = \frac{x_2-\De^2 \ve}{x_1 -\De \ve } = \frac{x_2 x_4 - x_5^2 }{ x_1 x_4 - x_5 x_3 } \\
% \vi = \frac{(x_1-\De \ve )^2 }{x_2-\De^2 \ve} = \frac{(x_1 x_4 - x_5 x_3 )^2 }{x_2 x_4 - x_5^2 }
% \end{eqnarray}
% but inconsitent in the sense that $\vi +\ve \neq  1$. \vk{can be made consistent: drop \eq{x2=}.  }
% \subsubsection{Three Compartments}
In the following derivation all parameters are restricted to be strictly positive. 
Let's express all unknowns in terms of $\vf$. From simple algebra applied to \eqs{x3=}{x4=} and then \eq{x5=} we find
\begin{eqnarray}
\Dr = \frac{x_4-\Df^2\vf}{x_3-\Df\vf}\,,\ \ve = \frac{(x_3-\Df\vf)^2}{x_4-\Df^2\vf}\,,\ \De = \frac{x_5}{x_3-\Df\vf} \label{eq:sol1}
\end{eqnarray}
The intra-axonal parameters are expressed from \eqs{x1=}{x2=}, 
\begin{eqnarray}
\Di = \frac{x_2-\De^2\ve}{x_1-\De\ve} = \frac{x_2 x_4 -x_5^2  - \Df^2 \vf x_2}{x_1 x_4 - x_3 x_5 + \Df \vf x_5 - \Df^2 \vf x_1} \label{eq:sol2}
\end{eqnarray}
and 
\begin{eqnarray}
\vi =  \frac{(x_1-\De\ve)^2}{x_2-\De^2\ve} = \frac{(x_1 x_4 - x_3 x_5 + \Df \vf x_5 - \Df^2 \vf x_1)^2}{(x_4 - \Df^2 \vf) (x_5^2 - x_2 x_4 + \Df^2 \vf x_2)} \label{eq:sol3}
\end{eqnarray}
Now, all expressions depend exclusively on the unknown $\vf$. The last equation $\vi+\ve+\vf=1$ solves for $\vf$ as follows
\begin{eqnarray}
1-\vf &=& \vi + \ve \\
% &=& \frac{(x_1 x_4 - x_3 x_5 + \Df \vf x_5 - \Df^2 \vf x_1)^2}{(x_4 - \Df^2 \vf) (x_5^2 - x_2 x_4 + \Df^2 \vf x_2)} + \frac{(x_3-\Df\vf)^2}{x_4-\Df^2\vf} \\
&=& \frac{\Df^2 \vf x_1^2 - x_1^2 x_4 - x_2 x_3^2 - \Df^2 \vf^2 x_2 + 2 x_1 x_3 x_5 + 2 \Df \vf x_2 x_3 - 2 \Df \vf x_1 x_5}{x_5^2 - x_2 x_4 + \Df^2 \vf x_2}
\end{eqnarray}
% wir haben (x_4 - w*Df^2 ) gekürzt
Multiplying both sides by the denominator $(x_5^2 - x_2 x_4 + \Df^2 \vf x_2)$ (which is allowed since $x_5^2 - x_2 x_4 + \Df^2 \vf x_2 = -\Di^2 \Dr^2 \ve \vi \neq 0$)
leads to the following equation in $\vf$ 
% \begin{eqnarray}
% (x_2 \Df^2 - \Df^2 x_1^2 + 2 \Df x_1 x_5 - 2 x_2 x_3 \Df - x_5^2 + x_2 x_4) \vf & \\ + (x_4 x_1^2 - 2 x_1 x_3 x_5 + x_2 x_3^2 + x_5^2 - x_2 x_4) &= 0
% \end{eqnarray}
\begin{equation}
(x_2 \Df^2 - \Df^2 x_1^2 + 2 \Df x_1 x_5 - 2 x_2 x_3 \Df - x_5^2 + x_2 x_4) \vf  + (x_4 x_1^2 - 2 x_1 x_3 x_5 + x_2 x_3^2 + x_5^2 - x_2 x_4) = 0   \label{eq:last}
\end{equation}
This equation is linear, since the quadratic terms in $\vf$ cancel, which results in the unique final solution 
\begin{eqnarray}
\vf = \frac{x_2 x_4 -x_2 x_3^2  - x_1^2 x_4 - x_5^2 + 2 x_1 x_3 x_5}{
x_2 x_4 + \Df^2 x_2 - x_5^2 - \Df^2 x_1^2 - 2 \Df x_2 x_3 + 2 \Df x_1 x_5} \label{eq:sol_vf}
\end{eqnarray}
By inserting this $\vf$ into \eqss{eq:sol1}{eq:sol3} we obtain the full solution for 
all parameters, which is our main result.

We now analyze the case of zero denominator in \eq{eq:sol_vf}, which
will express an ambiguity inherent to the model itself.
Substituting the defining expression for the $x_i$'s, \eqss{x1=}{x5=}, gives for the denominator the form 
$\ve \vi (\Di \Dr - \Di \Df + \De \Df)^2$ 
and the same form multiplied with $\vf$ for the numerator. 
This means that for the special case 
\begin{eqnarray}
\Di \Dr - \Di \Df + \De \Df = 0 \label{eqspecial}
\end{eqnarray}
there is no information about $\vf$, since \eq{eq:last} turns into an identity. In other words, the constraint $\vi+\ve+\vf=1$ is automatically fulfilled for any $\vf$. 
Note that the system of \eqss{x1=}{x5=} is linear in the volume fractions. 
In particular, given the diffusivities, \eqS{x1=}{x3=}{volfrac}  can be used to build a linear system for the volume fractions. 
The determinant of the so constructed system is just 
the left-hand side of \eq{eqspecial} -- its zero value implies a linear dependency, thus resulting in an infinite number of solutions.

\begin{figure}[t]
\centering
\includegraphics[width=12cm]{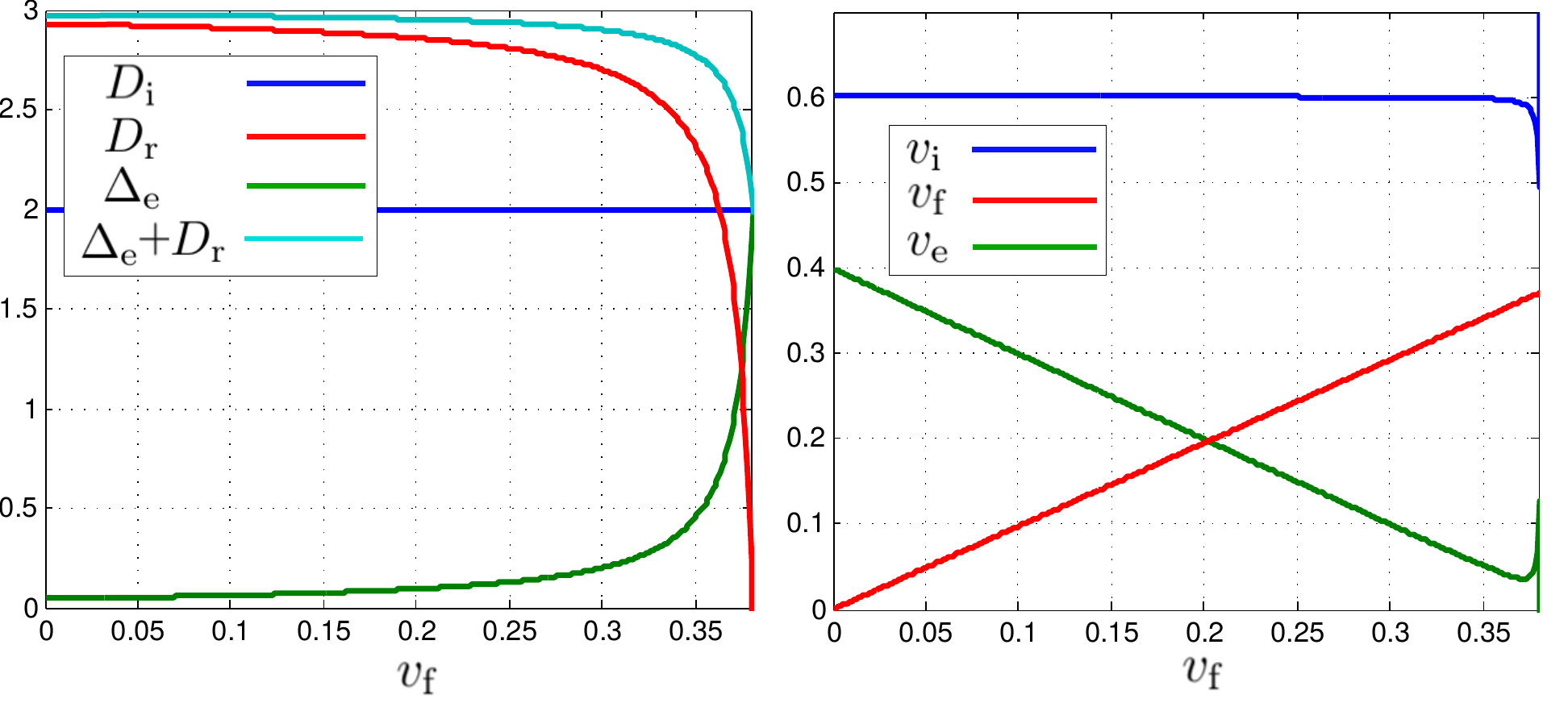}
\caption{An example for a family of solutions where $\Di \Dr - \Di \Df + \De \Df = 0$ for varying $\vf$. 
All these solutions have the same linear and planar moments up to order two. 
}
\label{fig:family}
\end{figure}

To understand the physics behind the degeneracy condition, \eq{eqspecial}, consider first two special cases. If $\Dr=0$, \eq{eqspecial} gives $\De=\Di$, which means that the extra-axonal compartment is indistinguishable from the intra-axonal one. Another case is the isotropic extra-axonal compartment, $\De=0$, in which case it is indistinguishable from free water, $\Dr=\Df$. We found a family of solutions that interpolates between 
these two special cases, which is shown in Figure \ref{fig:family}, as a functions of $\vf$. 
This solution only exists for 
a specific choice of diffusivities obeying \eq{eqspecial}. All the shown solutions
have exactly the same moments up to second order.
Outside the displayed interval, the solution is unphysical with several negative parameters. Interestingly, the intra-axonal diffusivity is not subjected to the ambiguity. 
In that case, one can find $\Di = \frac{\Df x_2}{\Df x_1 -x_5 }$.  

Note the similarity of the above degeneracy to the bi-exponential model when the diffusivities in two compartments are equal. \Eq{eqspecial} expresses this inherent drawback of multi-exponential models exemplified by the standard white matter model.

\subsection{Determination of mesoscopic dispersion $f_2$}

\Eq{f2=} operates with the moments of the order $b^2$. Here we show that the dispersion can also be expressed directly in terms of the signal.
Recall that the moments $W_\alpha^{l,k}$ define the Taylor expansion of $S_\alpha^{l}(b)$ in powers of $b$ according to \eq{W=}. Therefore the function 
\begin{eqnarray}
F(b) := -\frac{7}{4}\, \frac{S_\text{lin}^{2}(b) - 2 S_\text{pla}^{2}(b)}{S_\text{lin}^{0}(b) - S^0_\text{pla}(b)} 
\end{eqnarray}
reproduces \eq{f2=} with account for the identities $W_\text{lin}^{0,1} =  W_\text{pla}^{0,1} $ and $W_\text{lin}^{2,1} = 2 W_\text{pla}^{2,1}$. Practically, one has to consider the function 
%% Original equation:
% \[
% F(b) = f_2 + \alpha f_2 b + \mathcal{O}(b^2) = 
% f_2 -  \frac{f_2}{14} \,\frac{\vi \Di^3 + \ve \De^3}{\vi \Di^2 + \ve \De^2} b + \mathcal{O}(b^2)
% \]
\[
F(b) = f_2 + \mathcal{O}(b) \,
\]
and fit it linearly to find its value for $b=0$. 
%Account for the $b^2$ terms might help to avoid a bias (on the costs of precision). 
%\vk{Do we need a small numerical experiment concerning the fitting accuracy/precision?}

%\vk{skept to subsection 2.3}
\subsection{Linear and spherical encodings are not sufficient}
\begin{figure}[t]
\includegraphics[width=14cm]{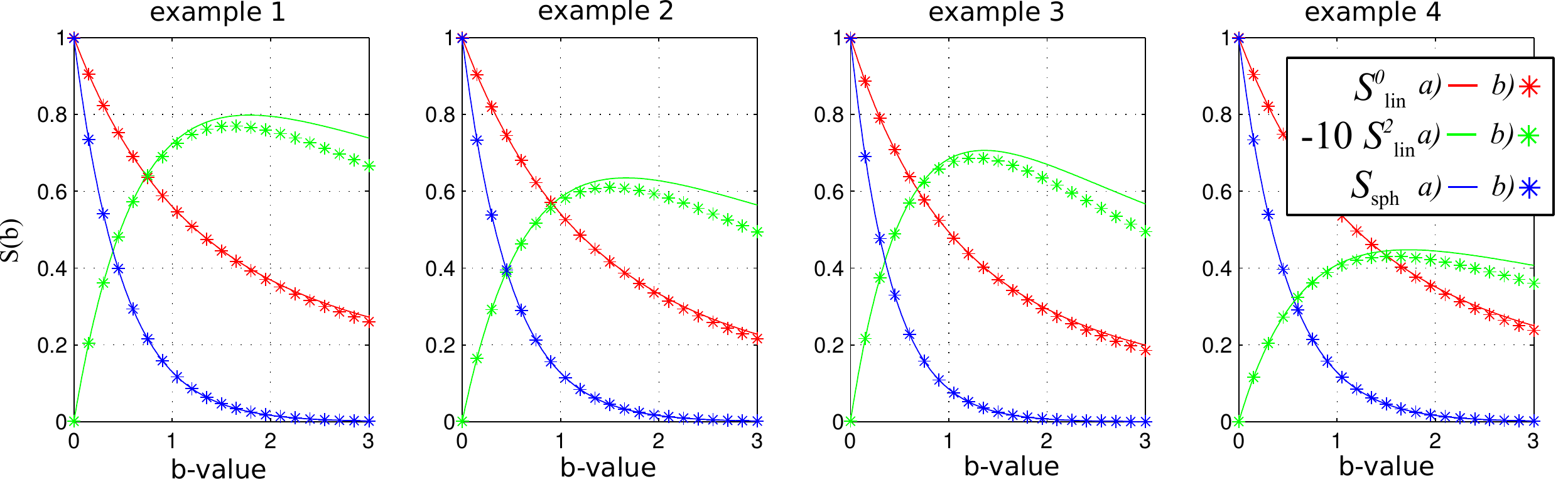}
\caption{Signal courses for the four examples, where linear and spherical moments are 
identical up to order two. First notable differences appear 
above $b=2$. Note that for $S^2_\text{lin}$ differences are enlarged 
by a factor of ten. 
}
\label{fig:example} 
\end{figure}

%\vk{I bet it can be made more convincing on the simple analytical level.} 
For spherical encoding we have
%\vk{define above: spherical = isotropic. }
\begin{eqnarray}
\Ms(b) &=& \vi e^{-b \Di/3} + \ve e^{-b (\De/3+\Dr)} + \vf e^{-\Df b} \\
W_\text{sph}^k &=& \left. \frac{d^k}{db^k} \right|_{b=0} \Ms(b)
\end{eqnarray}
We assume that only moments up to $\mathcal{O}(b^2)$ are observable, i.e.
$W_\text{lin}^{0,1},W_\text{lin}^{0,2},W_\text{lin}^{2,1},W_\text{lin}^{2,2},W^1_\text{sph},W^2_\text{sph}$
are known. We know that $W^1_\text{sph}$ is linearly dependent on $W_\text{lin}^{0,1}$
and $W_\text{lin}^{0,2}$, so linear and spherical encodings give five equations up to order 2. 
In fact, with these equations, one can find analytically a solution for the two-compartment 
model without the fast water fraction. However, this solution has two roots and is, hence, ambigious. 
We do not show here the solutions, but give a few numeric examples, where both roots lead to physical meaningful results:
\begin{center}
\begin{tabular}{l|l|l|l|l|l}
solution & $\Di$ & $\De$ & $\Dr$ & $f_2$ & $\vi$ \\ \hline 
1.a & 2.00 & 0.60 & 0.50 & 0.80 & 0.60 \\ 
1.b & 2.11 & 1.29 & 0.24 & 0.74 & 0.31 \\ \hline 
2.a & 2.00 & 0.60 & 0.50 & 0.80 & 0.40 \\ 
2.b & 2.17 & 1.11 & 0.31 & 0.72 & 0.17 \\ \hline 
3.a & 2.40 & 1.00 & 0.50 & 0.80 & 0.40 \\ 
3.b & 2.58 & 1.51 & 0.31 & 0.75 & 0.14 \\ \hline 
4.a & 2.00 & 0.60 & 0.50 & 0.50 & 0.50 \\ 
4.b & 2.14 & 1.20 & 0.27 & 0.46 & 0.24 \\ \hline 
\end{tabular}
\end{center}
where mainly $\vi,\De$ and $\Dr$ are confused. The parameters $\Di,f_2$ and
$\De+\Dr$
are rather stable. This goes in line with the observation that a spherical 
encoding can resolve the ambiguity of the parallel 
diffusivities \cite{fieremans2011white,fieremans2018, dhital2017absence}, but still has to struggle with $\De$, $\Dr$ 
and $\vi$.
In Figure \ref{fig:example} we show signal courses for the counterexamples.

\section{Conclusion}
We have constructively shown that linear and planar diffusion encodings can fully resolve the three-compartment model of white matter using 
data up to the order $\mathcal{O}(b^2)$ and $l=2$. 
The common experience with the diffusional kurtosis imaging \cite{jensen2005diffusional} indicates the practical availability of $\mathcal{O}(b^2)$ terms. While in principle, these terms include information for $l \leq 4$, the order $l=4$ is spoiled by noise as it was shown for a typical two-shell measurement on an advanced scanner with the maximal gradient strength $80\units{mT/m}$ \cite[Fig.\,2]{reisert2017disentangling}.  

% We further highlighted a special situation where the solution gets ambiguous due to an inherent confusion between extra-axonal water and free water,
% which shows that under certain circumstances prior knowledge is indeed necessary to make stable estimates. \vk{disagree: `inherent' means no cure possible. Do not use the model if so. In other words, in the case of degeneracy, the results are infinitely sensitive to the priors, without any influence of data. Suggest replacing this paragraph with the following one:}

Our analysis highlighted a special situation of ambiguous solution due to an inherent inability of multiexponential models to resolve compartments with similar parameters. 
The only way to distinguish such compartments is measuring in a domain where their differences get 
apparent, for example in the large b-value regime, where stable estimates of higher order information becomes possible. Without such information, a stable parameter estimate is only possible relying on prior knowledge.
%However, if large b-values are not accessible, prior knowledge \vk{;-)} has to be used to make stable estimates.

We have also shown that a combination of spherical and linear encoding is not 
enough to find a unique solution in order $\mathcal{O}(b^2)$. In fact,  
$\mathcal{O}(b^2)$ information delivered  by a spherical encoding is 
fully contained in the combination of linear and planar information,
namely $W_\text{sph}^{0,2}  =  (4W_\text{pla}^{0,2} - W_\text{lin}^{0,2})/3$, which
renders a spherical encoding in the presence of linear and planar encodings
in the low b-value regime superfluous.
In fact, it is a fortunate coincidence that $\mathcal{O}(b^2)$ information spanned by
linear and planar diffusion encoding (it is actually the
'full' encoding in $\mathcal{O}(b^2)$) is 6 dimensional (\eqss{weq1}{weq6}) and the parameter space 
of the three compartment white matter model has also 6 free parameters, \eq{model}. 

The derived mapping is only valid for noiseless signals, i.e., when the
signal is in the image of the modeling equation. For practical applications the obtainable signal-to-noise ratios are too low. 
A recent preprint \cite{coelho2018double} shows by numerical 
simulations that in a slightly simplified setting (two-compartments and Watson distribution) also in the noisy case the degeneracy is resolved.
The importance of 
the analytical solution lies in its justification for parameter estimators that rely on unimodal posterior distributions.
Additionally, the solution can give certain hints for the construction of such parameter estimators. In fact, the expression
of the parameters are all low-order rational functions of the moments (which are all linear projections of the signal).
This suggests to make a similar approach for the estimator (e.g. as found in 
\cite{reisert2017disentangling}), i.e.\ using functions, which are rational in linear combinations of the signal.

\bibliographystyle{apalike}
\bibliography{sample}

\end{document}